\documentclass[10pt,letterpaper,twocolumn]{article}
\usepackage{ol2}
\usepackage{varioref}
\usepackage{graphicx}
\usepackage[T1]{fontenc}
\usepackage{subfigure}
\usepackage{cite}
\usepackage{amsmath}
\usepackage{amsfonts}
\usepackage{stmaryrd}
\usepackage{psfrag}
\usepackage{tabularx}\newcolumntype{Y}{>{\centering\arraybackslash}X}
\graphicspath{{/home/image/eps/pmma/}}
\DeclareGraphicsExtensions{.eps,.EPS,.ps,.PS,.eps.gz}
\begin{document}

\twocolumn[
\title{Doubly resonant photonic antenna for single infrared quantum dot imaging at telecommunication wavelengths}

\author{Zhihua Xie$^1$, Miguel Suarez$^1$, Mathieu Mivelle$^{1,2}$, Roland Salut$^1$, Jean-Marc Merolla$^1$ and Thierry Grosjean$^1$*}

\address{$^1$Optics Department - FEMTO-ST Institute UMR 6174 - Univ. Bourgogne Franche-Comte - CNRS - Besancon, France}

\address{$^2$ Current address: Université Pierre et Marie Curie, CNRS, Institut des NanoSciences de Paris, UMR 7588, 75005 Paris, France}

\email{*thierry.grosjean@univ-fcomte.fr}

\begin{abstract}
Colloidal Quantum dots (CQDs) are nowadays one of the cornerstones of modern photonics as they have led to the emergence of new optoelectronic and biomedical technologies. However, the full characterization of these quantum emitters is currently restricted to the visible wavelengths and it remains a key challenge to optically probe single CQDs operating in the infrared spectral domain which is targeted by a growing number of applications. Here, we report the first experimental detection and imaging at room temperature of single infrared CQDs operating at telecommunication wavelengths. Imaging was done with a doubly resonant bowtie nano-aperture antenna (BNA) written at the end of a fiber nanoprobe, whose resonances spectrally fit the CQD absorption and emission wavelengths. Direct near-field characterization of PbS CQDs  reveal individual nanocrystals with a spatial resolution of 75 nm ($\lambda/20$) together with  their intrinsic 2D dipolar free-space emission properties and exciton dynamics (blinking phenomenon). Because the doubly resonant BNA is strongly transmissive at both the CQD absorption and emission wavelengths, we are able to perform all-fiber nano-imaging with a standard 20 \% efficiency InGaAs avalanche photodiode (APD). Detection efficiency is predicted to be 3000 fold larger than with a conventional circular aperture tip of the same transmission area. Double resonance BNA fiber probes thus offer the possibility of exploring extreme light-matter interaction in low band gap CQDs with current plug-and-play detection techniques, opening up new avenues in the fields of infrared light emitting devices, photodetectors, telecommunications, bio-imaging and quantum information technology.
\end{abstract}

\maketitle ]

\section{Introduction}

Nanotechnologies rely on the ability to fabricate and manipulate nanosize structures, and to detect these nano-elements individually, and characterize their intrinsic properties. The need for nanoscale detection and imaging is of crucial importance in the rapidly growing optoelectronic and biomedical technologies based on the engineering, manipulation and control of colloidal quantum dot (CQD) \cite{alivisatos:science96}. In that domain, light emission is a direct means to locate CQDs and reveal their intrinsic nanoscale properties and dynamics. If the characterization of CQDs in the visible spectral range has been successfully achieved with optical imaging techniques directly derived from single molecule detection and spectroscopy, the characterization of CQDs in the infrared spectral domain remains extraordinary difficult, preventing a growing number of devices and applications targeting this wavelength range to reach their full potential.

Infrared CQDs are indeed among potential key building blocks in the engineering of modern infrared light sources, photodiodes and solar cells \cite{rogach:small07,sargent:am05,kershaw:jstqe00,keuleyan:natphot11}. Owing to quantum size effect, infrared CQDs have shown exceptional spectral tunability up to mid-infrared domain, thus leading to totally new paradigms in a wide field of applications ranging from telecommunications to biological imaging and photovoltaics \cite{tang:am12,harrison:pac00,medintz:natmat05,kim:natbiotechnol04}. Given their high photostability, infrared CQDs are also suitable candidates as single photon sources at room temperature for fiber compatible quantum information processing at telecommunication wavelengths.

Detecting and probing infrared CQDs at room temperature down to individual elements is highly desirable as it would allow for improved functionality and performance of these nanocrystals in the applications mentioned above. However, the lack of high numerical aperture (NA) imaging optics at infrared frequencies remains a major problem impeding the study of infrared CQDs at the single emitter level. High NA objectives commonly used for single molecule detection are not corrected to operate in the infrared portion of the spectrum, leading to a drop of photon collection efficiency and resolution at these wavelengths due to chromatism of the optics initially optimized for visible frequencies. To circumvent these important losses and detect single infrared CQDs, one way involves implementing a highly efficient photon detector \cite{correa:nl12}. However, higher detection efficiency is so far made at the expense of a detector operating at low temperature, which dramatically weighs down the acquisition system. To date, no study has reported the detection of single infrared nanocrystals at telecommunication wavelengths ($\lambda > 1300$ nm) with far-field imaging approaches.

Scanning Near-field Optical microscopy (SNOM) provides alternative imaging systems free from bulky optics and capable of probing single fluorescent molecules and quantum dots with resolution well beyond the diffraction limit \cite{hosaka:jm01,matsuda:prl03}. These plug-and-play nano-imaging architectures rely on a circular nano-aperture positioned at the end of a double taper fiber tip \cite{saiki:apl96} which acts as a near-field interface between the nano-emitters and the optical fiber. Unfortunately, subwavelength circular nano-apertures operate beyond cutoff in their waveguide transmission process, which dramatically limits their optical throughput and collection efficiency, especially at near-infrared frequencies \cite{tsumori:ao11}.

The development of photonic nano-antennas has recently broken down the cutoff barrier in nanoscale optics, thus enabling unprecedented light enhancement and confinement down to the nanometer scale in a spectral domain spanning from visible to infrared regime (and beyond) \cite{novotny:natphot11}. Aperture nano-antennas, such as bowtie nano-aperture antenna (BNA)\cite{wang:nl06}, have been successfully integrated at the apex of fiber tips for nano-imaging applications \cite{mivelle:ox10}. Their nanoscale resonant transmission properties below cutoff led to background-free ultra-bright light sources for single molecule imaging at visible frequencies \cite{mivelle:nl12} and highly sensitive light nano-collectors for photonic crystal near-field mapping at the infrared telecommunication window \cite{vo:ox12}. In these applications, the  nano-antenna single resonance is tuned to the portion of the spectrum related to the study.

In this paper, we propose a new type of nano-antenna fiber probe based on a doubly resonant BNA for imaging single infrared CQDs at telecommunication wavelengths. Our design relies on a BNA showing two spectrally separated resonances tuned to the nanocrystal's absorption and emission wavelengths (visible and telecommunication spectra, respectively), thus providing an efficient two-way communication channel between single CQDs and the optical fiber for both nanocrystal optical excitation and fluorescence collection. This new approach is demonstrated with the first imaging of single PbS CQDs at telecommunication wavelengths ($\lambda$=1500 nm). The sensitivity of the resulting fiber probe is large enough to detect single CQDs with a conventional 20\% efficiency  InGaAs photon counter.  The resulting all-fiber optical bench is then ultracompact (free from bulky optics) and yields $\lambda/20$ spatial resolution (i.e. 75 nm). Beyond the imaging of the spatial distribution of single PbS CQDs on a surface, the new BNA nanoprobe reveals PbS nanocrystal's 2D dipolar emission properties and intrinsic blinking phenomenon. These results are highly relevant in the context of the growing interest in understanding and controlling fundamental properties of semiconductor nano-crystals dedicated to the infrared regime and more especially to telecommunication wavelengths.

\section{Tip simulation and fabrication}

\begin{figure}
\includegraphics[width=0.99\linewidth]{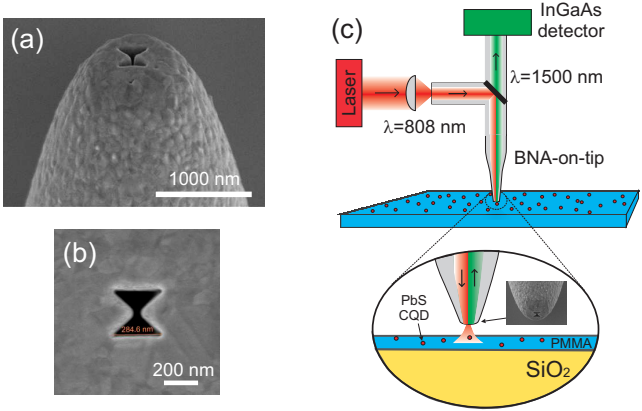}
\caption{(a,b) SEM micrographs of a doubly resonant BNA on a fiber tip: (a) side view of the tip apex and (b) top view of the BNA. (c) Schematic diagram of the experimental configuration: in-fiber excitation of the BNA is performed at $\lambda$=808 nm with light linearly polarized along the polarization axis of the BNA. The tip is set in contact to single PbS quantum dots embedded in a 20 nm thick PMMA layer and is raster scanned across the sample. During scanning, fluorescence signal at wavelengths around 1500 nm is collected by the BNA and detected through the fiber with a standard InGaAS APD.}
\label{fig:fab_principe}
\end{figure}

In order to produce the doubly resonant BNA fiber probe, polymer tips are first grown by photopolymerization at the cleaved end facet of a single mode fiber (SMF-28) \cite{bachelot:ao01}. The tips are about 40 microns long and have radius of curvature of 500 nm at their apex. Next, the probes are metal coated with a few nanometers thick titanium adhesion layer followed by a silver layer that is 300-450 nm thick. Silver is chosen for its high conductivity and minimum losses ensuring strong antenna effect. The tip apex is then polished by Focused Ion Beam (FIB) milling to reduce metal layer thickness on top to 200 nm (for each production batch, one sacrificial metal-coated tip is cut by FIB in order to estimate metal thickness at the tip apex and thus calibrate the polishing process). Finally, a 280 nm wide BNA with a square gap of 45 nm is opened at the tip apex by FIB milling. \ref{fig:fab_principe}(a) and (b) display SEM micrographs of a resulting all-fiber nano-imaging platform.

\begin{figure}
\includegraphics[width=0.99\linewidth]{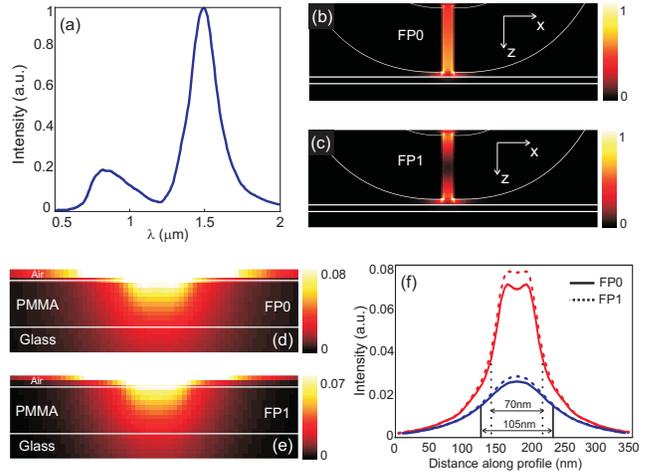}
\caption{(a) Theoretical resonance spectrum of a doubly resonant BNA on a tip, positioned at a distance of 15 nm beneath the surface of a 20 nm thick PMMA layer lying on a semi-infinite glass substrate. The excitation wave is launched into the tip body and is linearly polarized along the polarization axis (0x) of the BNA (ie. the symmetry axis of the BNA that crosses the two metal triangle). (c-f) Simulation of the optical electric intensity produced by (b,d) FP0 resonance at $\lambda=1500$ nm and (c,e) FP1 resonance at $\lambda=808$ nm in the longitudinal plane (0xz) that contains the BNA's polarization axis: (b,c) large scale views of the tip apex and (d,e) intensity plots right at the 20 nm thick PMMA layer. (f) Cross sections of the light spots shown in (d) (solid lines) and (e) (dashed lines), at the upper (red curves) and downer (blue curves) interfaces of the thin PMMA layer.}
\label{fig:simulation}
\end{figure}

The optical response of this new nano-antenna probe is simulated using three dimensional Finite Difference in Time Domain method (3D FDTD) available from a commercial software programme (Fullwave, Synopsis). \ref{fig:simulation}(a) shows the emission spectrum of a BNA probe placed 15 nm away from a 20 nm thick PMMA layer onto a semi-infinite medium of SiO$_2$.  The model used for all calculations of our nanoprobe consists of a volume spanning $\pm 3$ $\mu$m in ($0x$) and ($0y$) about the apex of the fiber tip. The apex, with a tip-radius of 500 nm, is located at x=y=z=0 and the simulation spans 1.5 $\mu$m below the tip in air, and terminates 5 $\mu$m into the body of the tip. The tip geometry considered in this study is that of a polymer tip of refractive index equal to 1.52 and taper angle of 16$^{\circ}$. Silver and PMMA are considered dispersive. The dispersion of silver is described by a Drude Lorentz model. The index of the glass substrate is set to 1.5. All six boundaries of the computation volume are terminated with perfectly matched layers in order to avoid parasitic unphysical reflections around the probe. The non-uniform grid resolution varies from 25 nm for portions at the periphery of the simulation, to 5 nm within the volume of the BNA and at the region immediately around the structure. Along ($0z$) axis, gridding is set down to 2 nm within the PMMA layer and 1.5 nm at the air gap between the tip and the sample. In the simulation, the BNA is excited with a gaussian beam launched within the tip at a distance of 4.5 microns from the BNA. The incident waves are temporally described by a single pulse and are linearly polarized along the BNA's symmetry axis that crosses its two metal triangles (parallel to (0x)). The time-varying electric field component $E_x$ transmitted by the nano-antenna is recorded at a single cell aligned with the tip symmetry axis and located 20 nm away from the BNA, within the PMMA layer (ie, 5 nm bellow the free interface of the PMMA layer). The spectrum of $|E_x|^2$ is calculated by Fourier-transforming this result.

The nanoprobe shows two resonances at wavelengths around 800 nm and 1500 nm, which correspond to the two first Fabry-Perot resonances of the BNA. The nano-structure enhances and confines light on the basis of the excitation of a highly confined guided mode bound to its nanometer size gap \cite{guo:ox08,ibrahim:ol10}. A BNA of finite thickness imposes boundary conditions to this guided mode and leads to Fabry-Perot resonances, which is a unique property  to all metallic nano-apertures operating below cutoff \cite{cao:prl02,liu:nature08}. The first resonance (at $\lambda=$1500 nm), noted FP0, is excited right at the cutoff of the nanoscale waveguide (it has near-zero effective refractive index) and thus does not undergo spectral dependence on BNA thickness. The second resonance (at 808 nm), noted FP1, is the first Fabry-Perot harmonic of finite phase velocity which can be spectrally tuned with the structure thickness. Such a response discrepancy between FP0 and FP1 with respect to antenna geometrical parameters allows for spectrally tuning the two BNA resonances independently from each other: FP0 spectral response is mainly governed by the BNA lateral size and gap width whereas FP1 is controlled by varying metal layer thickness (ie. aperture length) \cite{guo:ox08,ibrahim:ol10}. Both FP0 \cite{mivelle:ox10,vo:ox12} and FP1 \cite{mivelle:nl12} resonances of a BNA have demonstrated remarkable performances for near-field microscopy, in terms of resolution and throughput,  but have never been combined yet for optical application. Note that FP0 mode has demonstrated remarkable performances in extraordinary optical transmission in BNA arrays \cite{kinzel:ol10}and arrays of double resonance apertures have been used for enhancing second harmonic generation in thin layers of non-linear materials \cite{barakat:ox10}.

\ref{fig:simulation}(b-e) show simulations of electric intensity distributions produced by the BNA on a tip, at its two resonance wavelengths (b,d) 1500 nm and (c,e) 808 nm. The tip model is the same as the one used for spectrum calculations except that a continuous waves is launched into the tip at the two wavelengths of interest (one simulation per wavelength). Intensity plots are simulated along the longitudinal (0xz)-plane,  for an excitation polarization parallel to the BNA's polarization axis (0x). We see from \ref{fig:simulation}(b) and (c) that the two FP0 and FP1 modes are strongly bound to the BNA nanometer-scale feed gap: FP0 shows an almost uniform intensity distribution along the gap which evidences near-zero  effective refractive index of the mode, whereas FP1 is described by a single intensity node at half the BNA thickness. Both resonances allow for dramatically confining optical fields within the PMMA layer, as shown in \ref{fig:simulation}(d) and (e). The spot size within PMMA layer varies from 70 nm (better than $\lambda/20$) right underneath its top free surface up to 105 nm (around $\lambda/15$) at the bottom PMMA/glass interface (see \ref{fig:simulation}(f)).

To better evaluate the performances of the doubly resonant BNA on a tip, we compared our nanoprobe to conventional circular aperture tips in terms of detection efficiency. To this end, we numerically simulated the fraction of spontaneous emission from a single QE that is collected by the two types of nano-apertures, opened on the same tip, and in near-field interaction with the QE. Following a classical approach of the coupling between a QE and a nano-antenna, where the QE is modeled with a radiating dipole \cite{novotny:book,kaminski:jctn07}, the ratio of detection efficiencies $R$ between the two types of aperture tips used in emission/collection mode can be defined as:

\begin{equation}
\centering
R=\frac{|\textbf{n} \cdot \textbf{E}_{BNA}(\textbf{r})|^2}{|\textbf{n} \cdot \textbf{E}_{circ}(\textbf{r})|^2} \frac{q_{tip}^{BNA}}{q_{tip}^{circ}},
\label{eq:q_tip}
\end{equation}

where $\textbf{n}$ is the unit vector along the absorption dipole moment of the QE, and $\textbf{E}_{BNA}(\textbf{r})$ and $\textbf{E}_{circ} (\textbf{r})$ are the complex electric field vectors emanating from doubly resonant BNA and circular aperture, respectively, at the dipole's position $\textbf{r}$ and at excitation wavelength ($\lambda=$808 nm). $q_{tip}^{BNA}$ and $q_{tip}^{circ}$ are the collection yields within the BNA and circular aperture tips, respectively, calculated at QE emission wavelength ($\lambda=$1500 nm) as the power transmitted into the tips (via Poynting vector flux) normalized by the total power dissipated by the dipole source \cite{novotny:book,kaminski:jctn07}.  Tip and sample models remain unchanged, the two aperture tips are considered to probe a single quantum emitter of absorption and emission dipole moments oriented along (0x) and positioned along the tip symmetry axis and 5 nm underneath the free surface of the PMMA layer. The excitation fields at dipole's position are calculated in continuous wave regime in a way similar to that shown above for 2D field plots($\lambda$=808 nm). Collection yields are obtained by inserting a dipole source at the desired QE's position radiating a continuous wave at QE's emission wavelength ($\lambda$=1500 nm). We predict from \ref{eq:q_tip} that the fluorescence signal collected by the tip is 3000 times larger with the doubly resonant BNA than with a 226 nm wide circular aperture probe of identical transmission area. Note that the resolution ability of the resulting circular aperture is comparatively modest, emphasizing the potential of our doubly resonant BNA to probe single infrared CQDs.

\section{Results and discussion}

To demonstrate experimentally the doubly resonant BNA for single infrared CQD imaging, we probed single PbS quantum dots (from Evident) whose luminescence center wavelength is about 1500 nm. The choice of PbS CQDs is motivated by their high potentialities in photovoltaics and photodetectors \cite{macdonald:natmat05,zhao:nl10} as well as in the realization of tunable light sources, amplifiers and modulators at telecommunication wavelengths \cite{bakueva:apl03,steckel:am03,grandidier:nl09,klem:apl05}.

PbS CQDs  are randomly embedded in a 20 nm thick PMMA layer deposited onto a microscope coverslip. Preparation of QD samples. We used PbS quantum dots from Evident Company diluted at a concentration of $22.7$ nmol/mL.  $1.5$ $\mu$L of original solution is diluted in $85$ $\mu$L toluene and $15\mu$L PMMA (4\% PMMA in anisole). The mixture is kept  for $30$ minutes in an ultrasound bath to cancel QD aggregates. The solution is then spin coated onto a microscope coverslip at a speed of $2000$ rps. Prior to spin coating, the microscope coverslip is cleaned in a mixture of alcohol and acetone (held for $10$ minutes in an ultrasound bath) and dried by $N_{2}$ gas blowing. After spin coating, the QDs are randomly distributed within a 20 nm thick PMMA layer (layer thickness has been measured by Atomic Force Microscopy).

The doubly resonant BNA nanoprobe is mounted in a commercial SNOM (NTegra, from NT-MDT company) to be raster scanned across single CQDs at a nanometer distance from the sample surface. Tip-to-sample distance is kept constant (within 15 nm) with a shear-force feedback loop. Single CQDs are locally excited by the nano-antenna with in-fiber illumination at $\lambda$=808 nm from a laser diode (BlueSky Research). The impinging laser light is spectrally filtered before reaching the nano-antenna to avoid the detection of an undesired parasitic background signal at the nanocrystal's emission wavelengths. About 50 $\mu$W are injected into the fiber tip. The proper polarization state is also defined with a fiber polarization controller.  Spontaneous emission is collected by the same nano-antenna and transferred backward through the same fiber to an add-drop filter which extracts and transfers fluorescence photons towards 20 \% efficiency InGaAs photon counter (LYNXEA-M1 from AUREA Technology company). The detector is directly connected to the SNOM control unit for image acquisition.

\begin{figure}
\includegraphics[width=0.99\linewidth]{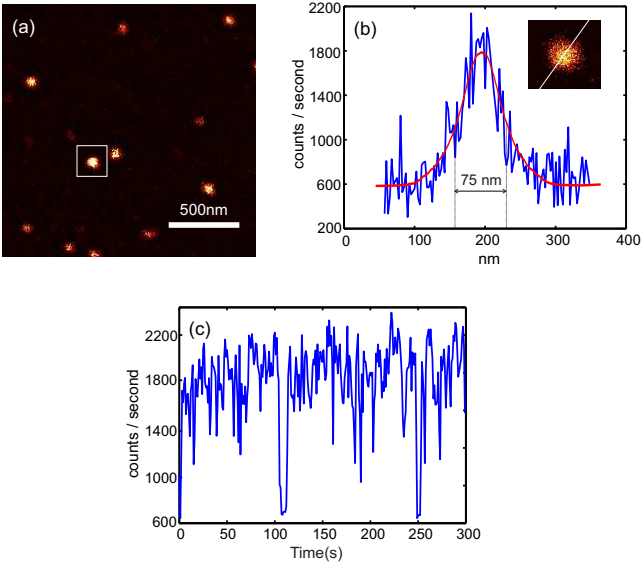}
\caption{The doubly resonant BNA on fiber tip reveals single near-infrared PbS CQDs as single tiny circular spots. (a) Image of the quantum dots with the all-fiber platform, the in-fiber excitation waves are polarized along the BNA polarization axis and the signal is collected through the same nanostructure. (b) Intensity profile of a single quantum dot shown in the figure inset (along the white line of the inset): the quantum dot is located in the rectangle of (a).  (c) Time trace acquired when the doubly resonant BNA is located right at a single nanocrystal, revealing blinking phenomenon (integration time: 1 second).} 
\label{fig:snomimage}
\end{figure}

The fluorescence image reported in \ref{fig:snomimage}(a) is 1.8 $\mu$m wide and is acquired at a scan speed of $2.34\mu$m/s. We see a series of single circular spots of various brightness and almost constant size. \ref{fig:snomimage}(b) displays a close range scan of one of the single quantum dots of \ref{fig:snomimage}(a). A Gaussian fit of the experimental cross section reveals a full width at half maximum at a single spot of 75 nm  ($\lambda/20$) which is in correct agreement with the spot size values predicted for both resonances of the BNA (\ref{fig:simulation}(f)). The signal-to-noise level, larger than 3, is high enough to localize the nanoscrystals onto the surface and analyze their spatial distribution and emission properties. \ref{fig:snomimage}(c) shows the time trace acquired when the nano-antenna probe is positioned directly on a single quantum dot. Two-state blinking is observed, confirming single quantum dot distribution within the PMMA layer. Because the input power at the BNA is kept at low level, photon emission is turned off during short time delays, which is in agreement with colloidal quantum dot emission properties \cite{banin:jcp99}. Note that the detector integration time chosen here reveals emission intermittency over periods larger than one second.
According to the simulation of the light confinements produced by the doubly resonant BNA within the PMMA layer (\ref{fig:simulation}(f)), the bright spots of almost identical FWHM close to 75 nm observed in \ref{fig:snomimage}(a) show that the quantum dots are mostly located near the upper interface of the PMMA layer.

Given the dipolar properties of BNAs \cite{jin:apl05,mivelle:nl12,vo:ox12}, \ref{fig:snomimage}(a) also provides information regarding the orientation properties of the transition dipole moment of single PbS CQDs. The distribution of near constant isotropic circular single-spot patterns observed within \ref{fig:snomimage}(a) do not match the near-field images of single fluorescent molecules which show a variety of different patterns under BNA imaging, depending on the orientation of the molecule's absorption dipole moment \cite{mivelle:nl12}. This confirms that the PbS CQD fluorescence process is not driven by a 1D oriented dipole moment but rather by a 2D dipole moment parallel to the sample surface, as is the case for visible colloidal quantum dots \cite{chung:proc03,empedocles:nature99}. They do not undergo a fixed transition dipole moment in the sample plane (i.e. they are not selective with respect to incident polarization), as for single molecules, and uniform isotropic optical patterns they are detected at nanocrystal positions despite the highly anisotropic BNA's optical response \cite{mivelle:nl12,vo:ox12}. This means that the orientation of the absorption and emission dipole moments are controlled by the two dipolar resonances of the BNA, respectively, leading to nearly constant single spot shapes within the acquisition area.  The different intensities measured throughout the single spot distribution may then reveal slight anisotropies in the 2D dipole moments of the CDQs (leading to more or less overlap with the BNA's dipole emission and collection) or quantum dots of non-uniform absorption cross-sections and/or intrinsic quantum yields.

\begin{figure}
\includegraphics[width=0.65\linewidth]{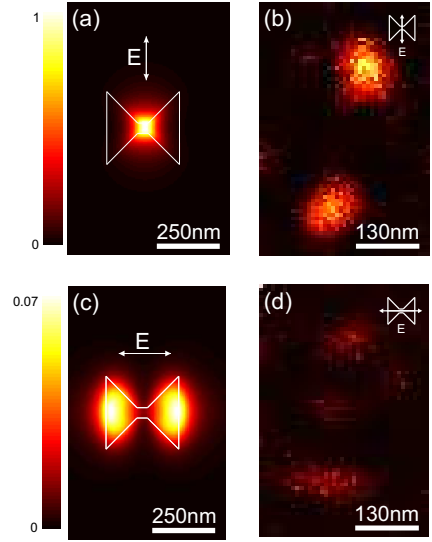}
\caption{The BNA is demonstrated in the local imaging of single PbS CQDs.  (a,c) Simulation of the electric intensity produced by the FP1 resonance of the BNA at $\lambda$=808 nm plotted in the transverse plane taken at half the thickness of the PMMA layer right at the feed gap of the BNA, for (a) and (c) two excitation polarization directions parallel and perpendicular to the BNA's polarization axis (see arrows), respectively. (b,d) Images of two quantum dots acquired for excitation polarization parallel and perpendicular to the BNA's polarization axis (see insets), respectively. These two images refer to the intensity distributions of FP1 resonance shown in (a) and (c), respectively.}
\label{fig:polar}
\end{figure}

The dipolar properties of the BNA, which are common to all gap based nano-antennas, associates the generation of a tiny "hot spot" to a high sensitivity of the nanostructure to the excitation polarization \cite{jin:apl05}: the tight optical confinement generated at antenna gap for polarization parallel to its polarization axis (\ref{fig:polar}(a)) is canceled and the intensity maximum over the antenna is greatly reduced if the polarization direction is rotated by 90$^{\circ}$ (\ref{fig:polar}(c)). \ref{fig:polar}(b) and (d) show the single PbS CQD response for excitation polarization parallel and perpendicular to the BNA's polarization axis, respectively. When the BNA is resonantly excited, the image shows well-defined spots beyond noise level, showing a FWHM about 75 nm. When the polarization direction is turned by 90$^{\circ}$, the single quantum dot patterns disappear and a weaker signal is detected, which is in qualitative agreement with the BNA responses shown in \ref{fig:polar}(a) and (c). These results prove that the single infrared CQD imaging process relies on the BNA itself.

\section{Conclusion}

In conclusion, we demonstrated the unprecedented ability of photonic antennas to bridge the gap between optical imaging and infrared CQD analysis at telecommunication wavelengths which remains relatively unexplored optically despite highly promising perspectives. We introduced the concept of doubly resonant BNA on a fiber tip as an ultracompact near-field optical probe enabling PbS CQD detection and imaging at a wavelength of 1.5 $\mu$m with a standard 20\% efficiency InGaAs photon counter.  The method is totally new and paves the way towards full optical characterization schemes of infrared CQDs down to single elements, leading to potential improvements in the development of devices and methods within the telecommunication spectral range. Our nanoprobe has great potential in the study, quantification and optimization of the antibunching effect from infrared CQDs capable of providing room temperature single photon sources at telecommunication wavelengths. Our all-fiber nano-antenna platform also holds high promises in harnessing these size tunable on-demand single photon sources to optical fiber networks for the development of ultracompact fibered architectures for quantum information processing at telecommunication wavelengths.
\\

The authors are indebted to Ulrich Fisher for useful discussions and Cara Leopold for her support in the paper writing. This work is funded by the Labex ACTION (ANR-11-LABX-0001-01).

\end{document}